\documentclass[11pt]{article}
\usepackage[utf8]{inputenc}

\usepackage{amssymb}
\usepackage[longnamesfirst]{natbib}
\usepackage{enumitem}
\setlist[itemize]{nosep,topsep=3pt,itemsep=3pt}
\setlist[enumerate]{nosep,topsep=3pt,itemsep=3pt}

\usepackage[backref=page]{hyperref}

\newif\ifbackrefshowonlyfirst
\backrefshowonlyfirstfalse
\makeatletter
\let\BR@direct@old@hyper@natlinkstart\hyper@natlinkstart
\renewcommand*{\hyper@natlinkstart}{\phantomsection\BR@direct@old@hyper@natlinkstart}
\let\BR@direct@oldBR@citex\BR@citex
\renewcommand*{\BR@citex}{\phantomsection\BR@direct@oldBR@citex}%

\long\def\hyper@page@BR@direct@ref#1#2#3{\hyperlink{#3}{#1}}

\ifx\backrefxxx\hyper@page@backref
    \let\backrefxxx\hyper@page@BR@direct@ref
    \ifbackrefshowonlyfirst
    \fi
\else
    \ifbackrefshowonlyfirst
    \fi
\fi

\RequirePackage{etoolbox}
\patchcmd{\Hy@backout}{Doc-Start}{\@currentHref}{}{\errmessage{I can't seem to patch backref}}
\makeatother

\usepackage{etoolbox}

\makeatletter

\patchcmd{\NAT@citex}
  {\@citea\NAT@hyper@{%
     \NAT@nmfmt{\NAT@nm}%
     \hyper@natlinkbreak{\NAT@aysep\NAT@spacechar}{\@citeb\@extra@b@citeb}%
     \NAT@date}}
  {\@citea\NAT@nmfmt{\NAT@nm}%
   \NAT@aysep\NAT@spacechar\NAT@hyper@{\NAT@date}}{}{}

\patchcmd{\NAT@citex}
  {\@citea\NAT@hyper@{%
     \NAT@nmfmt{\NAT@nm}%
     \hyper@natlinkbreak{\NAT@spacechar\NAT@@open\if*#1*\else#1\NAT@spacechar\fi}%
       {\@citeb\@extra@b@citeb}%
     \NAT@date}}
  {\@citea\NAT@nmfmt{\NAT@nm}%
   \NAT@spacechar\NAT@@open\if*#1*\else#1\NAT@spacechar\fi\NAT@hyper@{\NAT@date}}
  {}{}

\makeatother

\makeatletter
\newenvironment{pf}[1][\proofname]{\par
  \pushQED{\qed}%
  \normalfont \topsep0\p@\relax
  \trivlist
  \item[\hskip\labelsep\itshape
  #1\@addpunct{.}]\ignorespaces
}{%
  \popQED\endtrivlist\@endpefalse
}
\makeatother

\usepackage{amsthm,amsfonts,framed}
\usepackage[margin=1in]{geometry}
\usepackage{physics, graphicx}

\newcommand{\qaoa}{\text{QAOA}}
\usepackage[compact]{titlesec}
\titlespacing{\section}{0pt}{1.5ex}{0ex}
\titlespacing{\subsection}{0pt}{1.5ex}{0ex}
\titlespacing{\subsubsection}{0pt}{1ex}{0ex}
\titlespacing{\paragraph}{0pt}{1.5ex}{1ex}
\titleformat*{\paragraph}{\bfseries}

\usepackage{slantsc}
\theoremstyle{definition}
\newtheorem{definition}{Definition} 

\theoremstyle{plain}
\newtheorem{thm}{Theorem} 

\theoremstyle{plain}
\newtheorem{claim}{Claim}

\usepackage{tikz,xcolor,hyperref}
\definecolor{lime}{HTML}{A6CE39}
\DeclareRobustCommand{\orcidicon}{
	\begin{tikzpicture}
	\draw[lime, fill=lime] (0,0) 
	circle [radius=0.16] 
	node[white] {{\fontfamily{qag}\selectfont \tiny ID}};
	\draw[white, fill=white] (-0.0625,0.095) 
	circle [radius=0.007];
	\end{tikzpicture}
	\hspace{-2mm}
}
\foreach \x in {A, ..., Z}{\expandafter\xdef\csname orcid\x\endcsname{\noexpand\href{https://orcid.org/\csname orcidauthor\x\endcsname}
			{\noexpand\orcidicon}}
}

\hypersetup{
  colorlinks   = true,    
  urlcolor     = blue,    
  linkcolor    = blue,    
  citecolor    = blue 
}

\usepackage{xargs}
\usepackage[colorinlistoftodos,prependcaption,textsize=tiny]{todonotes}
\newcommandx{\bbnote}[2][1=]{\todo[linecolor=red,backgroundcolor=red!25,bordercolor=red,#1]{BB: #2}}
\newcommandx{\knote}[2][1=]{\todo[linecolor=blue,backgroundcolor=blue!25,bordercolor=blue,#1]{K: #2}}
\newcommandx{\bnnote}[2][1=]{\todo[linecolor=OliveGreen,backgroundcolor=OliveGreen!25,bordercolor=OliveGreen,#1]{BN: #2}}
\newcommandx{\improvement}[2][1=]{\todo[linecolor=Plum,backgroundcolor=Plum!25,bordercolor=Plum,#1]{#2}}
\newcommandx{\thiswillnotshow}[2][1=]{\todo[disable,#1]{#2}}

\DeclareMathOperator{\E}{\mathbb{E}}
\DeclareMathOperator\sgn{sgn}

\begin{document}

\title{Classical algorithms and quantum limitations for maximum cut on high-girth graphs}
\author{\orcidB{}Boaz Barak\thanks{Harvard University, \texttt{b@boazbarak.org}. Supported by NSF award CCF 1565264, a Simons Investigator Fellowship, and DARPA grant W911NF2010021.} \and \orcidA{}Kunal Marwaha\thanks{Berkeley Center for Quantum Information and Computation, \texttt{marwahaha@berkeley.edu}}  
}
\maketitle

\thispagestyle{empty}
\setcounter{page}{0}

\begin{abstract}
We study the performance of local quantum algorithms such as the Quantum Approximate Optimization Algorithm (QAOA) for the maximum cut problem, and their relationship to that of classical algorithms.

\begin{enumerate}

    \item We prove that every (quantum or classical) one-local algorithm (where the value of a vertex only depends on its and its neighbors' state) achieves on $D$-regular graphs of girth $> 5$ a maximum cut of at most $1/2 + C/\sqrt{D}$ for $C=1/\sqrt{2} \approx  0.7071$. This is the first such result showing that one-local algorithms achieve a value that is bounded away from the true optimum for random graphs, which is $1/2 + P_*/\sqrt{D} + o(1/\sqrt{D})$ for $P_* \approx 0.7632$~\citep{dembo2017extremal}.
    
    \item We show that there is a classical $k$-local algorithm that achieves a value of $1/2 + C/\sqrt{D} - O(1/\sqrt{k})$ for $D$-regular graphs of girth $> 2k+1$, where $C = 2/\pi \approx 0.6366$. This is an algorithmic version of the existential bound of \citet{lyons2017factors} and is related to the algorithm of   \citet*{Aizenman1987} (ALR) for the Sherrington-Kirkpatrick model. This bound is better than that achieved by the one-local and two-local versions of QAOA on high-girth graphs~\citep{hastings2019classical,Marwaha2021localclassicalmax}.
    
    \item Through computational experiments, we give evidence that the ALR algorithm achieves better performance than constant-locality QAOA for random $D$-regular graphs, as well as other natural instances, including graphs that do have short cycles.
    
\end{enumerate}

While our theoretical bounds require the locality and girth assumptions, our experimental work suggests that it could be possible to extend them  beyond these constraints. This points at the tantalizing possibility that $O(1)$-local quantum maximum-cut algorithms might be \emph{pointwise dominated} by polynomial-time classical algorithms, in the sense that there is a classical algorithm outputting  cuts   of equal or better quality \emph{on every possible instance}. This is in contrast to the evidence that polynomial-time algorithms cannot simulate the probability distributions induced by local quantum algorithms. 
\end{abstract}

\newpage
\section{Introduction}
Recent years have seen exciting progress in the construction of noisy intermediate-scale quantum (NISQ) devices \shortcites{Preskill_2018,bharti2021noisy} \citep{Preskill_2018,bharti2021noisy}.
One way to describe these devices is that they can implement the model of quantum circuits, but with the restriction that all gates respect a certain topology of a given graph $G$ (e.g., the qubits are associated with the vertices of the graph, and gates operate on either a single vertex or two neighboring vertices) and each operation involves a certain level of noise.
Due to the noise, computations on NISQ devices are inherently restricted to small \emph{depth}.
However, there is theoretical and empirical evidence that even at constant depth, such quantum circuits induce probability distributions that cannot be efficiently sampled by classical algorithms (see Section~\ref{sec:related} below).

While there is  evidence that NISQ devices could potentially achieve so-called ``quantum advantage'' (i.e., exponential speedup) for \emph{sampling problems}, the corresponding question for \emph{optimization problems} remains open. In particular, it is not known whether for natural optimization problems on graphs, a local quantum algorithm (a constant depth algorithm that in each step only operates on neighboring vertices)  can obtain better results than those achievable by polynomial-time classical algorithms, at least on some instances. 

A particular algorithm of interest is the \emph{Quantum Approximate Optimization Algorithm (QAOA)} \citep{farhi2014quantum}. The QAOA is parameterized by an integer $p$ and hyperparameters $\gamma_1,\ldots,\gamma_p$ and $\beta_1,\ldots,\beta_p$. For every $p$, $\qaoa_p$ for the maximum cut problem can be computed by a sequence of $p$ local unitaries (each acting only along edges), and so it is  \emph{$p$-local}, in the sense that for every vertex $v$, the output corresponding to $v$ depends only on the initial states of the vertices that are of distance at most $p$ from $v$ in the graph (see Section~\ref{sec:local} and Appendix~\ref{app:local} for more formal definitions).
\citet{farhi2014quantum} envisioned $p$ as an absolute constant not growing with $n$ (in which case the $\gamma_i$'s and $\beta_i$'s can be hardwired constants) or at worst growing very slowly with $n$.
Much of the excitement about QAOA is because for small values of $p$, $\qaoa_p$ can be (and in fact has been) implemented on near term devices \shortcites{Zhou_2020,Harrigan2021} \citep{Zhou_2020,Harrigan2021}. For example, \citet{Harrigan2021} implemented QAOA both for maximum cut and finding the ground state of the Sherrington-Kirkpatrick Hamiltonian. In both cases,  performance was maximized at $\qaoa_3$ since for larger $p$ the noise overwhelmed the signal.
Under widely believed complexity assumptions, we do not expect QAOA to solve maximum cut optimally in polynomial-time, or even beat the best classical approximation ratio in the \emph{worst case}.\footnote{Concretely, if the unique games conjecture is true and (as widely believed) $NP \nsubseteq BQP$, no quantum polynomial-time algorithm can obtain a better approximation ratio than \citet{goemans1995improved}'s classical algorithm   \citep{Khot02,KhotKMO07,mossel2010noise}.}
However, it is still very interesting to know whether there is some family $\mathcal{G}$ of graphs and some $p=O(1)$, on which $\qaoa_p$ or any other $p$-local quantum algorithm achieves exponential advantage over all classical maximum-cut algorithms.\footnote{Since maximum cut is NP-hard to approximate \citep{haastad2001some}, we can reduce the \emph{factoring} problem to approximating maximum cut on some family $\mathcal{G}$ of instances (namely the family resulting from this reduction). Hence under the assumption that factoring is hard, there exists some quantum polynomial-time algorithm that can approximate maximum cut on some family $\mathcal{G}$ better than all efficient classical algorithms. However this algorithm (which is based on \citet{shor1999polynomial}) will not be local and as far as we know cannot be implemented on NISQ devices.  \label{fn:factoring} }

\subsection{Our results}

In this work, we study the power and limitations of quantum and classical local algorithms for the maximum cut problem.
On input a graph $G=(V,E)$, an algorithm $A$ for the maximum cut problem outputs a vector $x\in \{ \pm 1 \}^V$, and the value of the cut $x$, denoted by $val(x)$, is the probability over $(i,j)\in E$ that $x_i \neq x_j$. 
We defer the formal definitions to Section~\ref{sec:local} and Appendix~\ref{app:local}, but roughly speaking, $A$ is \emph{$r$-local} if it begins by assigning some state to each vertex $v$, and for every vertex $u$, the final value $X_u$ depends only on the states of the vertices $v$ that are of distance at most $r$ from $u$ in the graph. 

Surprisingly, the following question is still open:

\paragraph{Question:} Do classical polynomial-time algorithms \emph{pointwise dominate} local quantum algorithms for maximum cut? In other words, is it true that for every $O(1)$-local quantum algorithm $A$ and $\epsilon>0$ there exists a polynomial-time algorithm $B$ such that \emph{for every graph} $G$, $val(B(G)) \geq val(A(G)) -\epsilon$?

\medskip We make some progress on this question by giving a positive and negative result for polynomial-time classical algorithm and local quantum algorithms respectively. We are not able to analyze either on every graph, but rather restrict ourselves to a (still exponentially large) family of instances:  all regular graphs of sufficiently high girth. 

In an $O(1)$-local algorithm, for every edge $\{ u,v \}$, the probability that $\{ u,v\}$ is cut only depends on a constant-radius ball around $\{u,v\}$. In $D$-regular graphs of sufficiently high girth all the neighborhoods (balls around a vertex of some distance $k$ sufficiently smaller than the girth) are isomorphic to the $D$-regular tree truncated at depth $k$. Hence in this case for every $O(1)$-local algorithm $A$, the probability an edge $\{ u,v \}$ is cut (and hence the expected value of the output cut) is equal to some value $f_A(D)$ that only depends on the algorithm $A$ and  degree $D$, and does not depend on the particular edge $\{u,v\}$ or any other details of the graph beyond the fact that its girth is sufficiently larger than the algorithm's locality. 
Hence, finding the best $k$-local algorithm for maximum cut amounts to finding the algorithm $A$ which maximizes $f_A(D)$. 

The value of $f_A(D)$ can be shown to be at most  $1/2 + O(1/\sqrt{D})$, because there exist high-girth graphs where this is the true optimum. In particular, by an eigenvalue bound one can show that the maximum cut of a random $D$-regular graph (which can be modified to have high girth) is at most $1/2 + 1/\sqrt{D}+o(1/\sqrt{D})$. Using a much more sophisticated argument, \citet{dembo2017extremal} showed  that the maximum cut of such graphs is in fact  $1/2 + P_*/\sqrt{D} \pm o(1/\sqrt{D})$ for $P_* \approx 0.763$ (see discussion below).

\begin{figure}[htbp]
    \centering
    \includegraphics[width=6in]{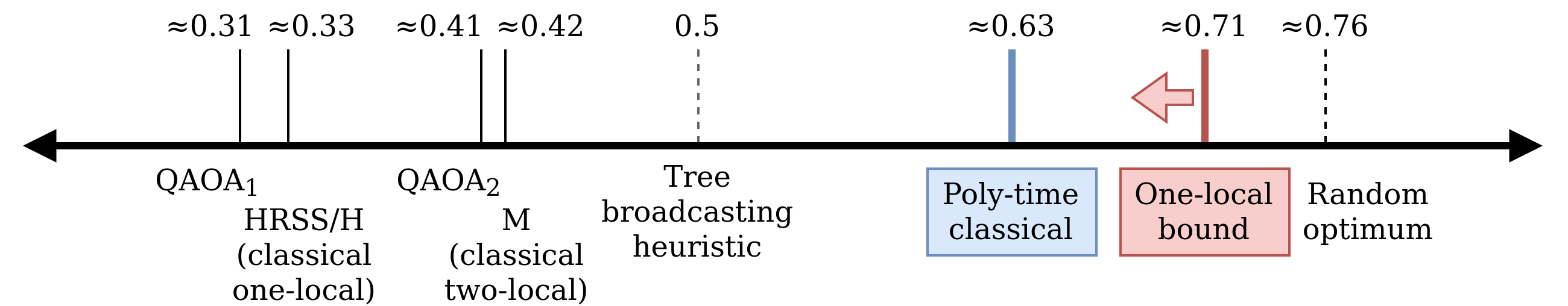}
    \caption{\footnotesize Upper and lower bounds on the value $C$ for which algorithms can guarantee a cut of value $1/2 + C/\sqrt{D}$ for $D$-regular graphs. Numerical analysis of $\qaoa_1$ and the one-local algorithm of  \citet{hirvonen2017large} was given by \citet{hastings2019classical}. Numerical analysis of $\qaoa_2$ and a two-local variant of the classical algorithm was given by \citet{Marwaha2021localclassicalmax}. Bounds on the maximum cut for random $D$-regular graphs were given by \citet{dembo2017extremal}. The point $C=1/2$ is interesting due to  a heuristic relation to the tree broadcasting model of \citet{evans2000broadcasting}. Our positive result for a polynomial-time (and $O(1)$-local) classical algorithm is marked in blue, while our negative result for one-local quantum and classical algorithms is marked in red.}
    \label{fig:results-overview}
\end{figure}

On the other hand, \citet{shearer1992note} gave a simple one-local classical algorithm that achieves at least $1/2 + C/\sqrt{D}$ for $C = \tfrac{2}{8} \approx 0.177$ on triangle free graphs, with improvements in the constant by \citep{hirvonen2017large,hastings2019classical,Marwaha2021localclassicalmax}, see Figure~\ref{fig:results-overview}.
We study the maximum value of the constant $C$ achievable by either classical polynomial-time algorithms or local quantum algorithms. We give a positive result (i.e., lower bound on $C$) for the former, and a negative result (i.e., upper bound on $C$) for the latter.

\paragraph{Classical algorithm for maximum cut on high-girth graphs.} We show that there is a classical polynomial-time algorithm that achieves a cut value  of $\approx 1/2 + 0.6366/\sqrt{D}$ on $D$-regular high-girth graphs:

\begin{thm}[Classical algorithm for maximum cut on high-girth graphs] \label{thm:algorithm-informal} There exists a polynomial-time classical algorithm $A$ that on input a $D$-regular graph $G$ of girth at least $g$, $A(G)$ outputs a cut of value at least
\[
\tfrac{1}{2} + \tfrac{C}{\sqrt{D}} - O\left(\tfrac{1}{\sqrt{g}} \right) \;,
\]
where $C = 2/\pi \approx 0.6366$.
\end{thm}

Prior to this work \citet*{hirvonen2017large} gave an algorithm that achieved a cut value of $1/2 + 0.28125/\sqrt{D}$ on $D$-regular triangle-free graphs. \citet{Marwaha2021localclassicalmax}, building on  \citep{hastings2019classical}, gave numerical evidence that there is a classical algorithm  achieving a cut value of $1/2 + 0.42/\sqrt{D}$ for $D$-regular graphs of girth larger than $5$.

Theorem~\ref{thm:algorithm-informal} is an algorithmic version of the existential bound by \citet{lyons2017factors}, who proved that for every sequence of graphs with girth tending to infinity, their maximum cut will  (in the limit) be at least $\tfrac{1}{2} + \tfrac{2}{\pi\sqrt{D}}$. While \citet{lyons2017factors}'s bound is existential, our approach is similar, and arguably makes explicit an algorithm that is implicit in his work. Our algorithm can also be thought of as a variant of the ALR algorithm (of \citet*{Aizenman1987}) for the Sherrington-Kirkpatrick (SK) Hamiltonian (c.f., \citet{panchenko2013sherrington}).
The SK model turns out to be closely related to maximum cut on random graphs.  
\citet*{dembo2017extremal} proved the maximum cut in random $D$-regular graphs (for $D$ tending to infinity) tends to $\tfrac{1}{2} + \tfrac{P_*}{\sqrt{D}}$  where $P_* \approx 0.7632$ is the ground state energy of the SK model.
For the SK model, the existential bound was made algorithmic by \citet{montanari2021optimization}. 
We conjecture that our algorithm is not optimal and that, just like the case of SK, the existential bound for maximum cut can can be made algorithmic, in the sense that for every $\epsilon>0$ there is a polynomial-time classical algorithm that on sufficiently high-girth $D$-regular graphs  outputs  a cut value of at least  $\tfrac{1}{2} + \tfrac{P_*}{\sqrt{D}} - \epsilon$. If this conjecture is true, then this classical algorithm would match or beat all $O(1)$-local classical or quantum algorithms on high-girth regular graphs, and (given Theorem~\ref{thm:limit-informal} below) would strictly outperform one-local quantum or classical algorithms.

\paragraph{Limitations for local quantum and classical algorithms.} 
While some limitations for $\qaoa_1$ and $\qaoa_2$ are already known (see Section~\ref{sec:related}), it is locality that makes QAOA suitable for NISQ devices. Therefore, it is important to study the limitations of more general quantum local algorithms. We make a first step in this direction, by showing that  (quantum or classical) one-local algorithms cannot achieve the maximum cut value for random $D$-regular graphs:

\begin{thm}[Limitations for one-local algorithms, informal] \label{thm:limit-informal} Let $A$ be a one-local (quantum or classical) algorithm. Then for every $D$-regular graph $G$ of girth at least $6$, the cut value output by $A$ is at most 
\[
\tfrac{1}{2} + \tfrac{C}{\sqrt{D}} \;,
\]
where $C=1/\sqrt{2} \approx 0.7071$.
\end{thm}

We defer the definition of locality to Section \ref{sec:local}. To the best of our knowledge, prior to this work, even for \emph{classical} one-local algorithms, the possibility of achieving the value  $1/2 + P_*/\sqrt{D}$ was not ruled out.  Figure~\ref{fig:results-overview} contains an overview of our results.

\paragraph{Our techniques.} Our positive result (classical algorithm) uses similar ideas to the prior works of \citet{lyons2017factors} and \citet{csoka2015invariant}. Our negative result (lower bound for one-local quantum or classical algorithms) is more original, and uses a technique that is qualitatively different than that of prior works; see Section~\ref{sec:related} for more discussion.

\paragraph{Local algorithms and tree broadcasting.} A $k$-local algorithm achieving $1/2 + C/\sqrt{D}$ value for maximum cut needs to satisfy two competing conditions. On the one hand, on average, every vertex has correlation $\rho = -2C/\sqrt{D}$ with its neighbor. On the other hand, locality of the algorithm means that the output of vertices that are sufficiently far apart in the graph are independent. A heuristic approach might be to assume that within the neighborhood of a vertex $u$, the probability distribution look as follows, $X_u$ is chosen uniformly from $\{ \pm 1 \}$, for every neighbor $v$ of $u$, $X_v=X_u$ with probability $1/2 - C/\sqrt{D}$ and $X_v \neq X_u$ with probability $1/2 + C/\sqrt{D}$, and these choices are done independently for each neighbor, and neighbor of neighbor, etc.  This process is known as the \emph{tree broadcasting process} \citep{evans2000broadcasting}, and it is known that long-range correlations exist if and only if $C> 0.5/\sqrt{D}$. Hence this heuristic approach might suggest that local algorithms would not be able to achieve values of $C$ larger than $0.5$. This turns out to be false, as our algorithm of Theorem~\ref{thm:algorithm-informal} is in fact $k$-local, though its locality does need to grow with the degree. It is still open whether or not $k$-local algorithms for $k \ll D$ can beat the value $C=0.5$.

\paragraph{Computational experiments.} Both our negative and positive results are restricted to high-girth graphs, and our negative result is further restricted in the sense that it only holds for one-local algorithms. However, through experiments, we demonstrate that even for larger $p$, for sufficiently large graphs, $\qaoa_p$ is dominated by the ALR algorithm on interesting families of graphs. These include both random $3$-regular graphs, as well as the \emph{torus} and \emph{grid} graphs, which are natural examples of graphs with short cycles. See discussion in Section~\ref{sec:experiments}, Figures \ref{fig:random}--\ref{fig:union}, and the Jupyter notebook at \url{http://tiny.cc/QAOAvsALR}.

\subsection{Related works} \label{sec:related}

Some bounds on QAOA's performance on the maximum cut problem high-girth graphs were given by  \citet{hastings2019classical} and \citet{Marwaha2021localclassicalmax}. Specifically \citet{hastings2019classical} showed that the $\qaoa_1$ algorithm achieves (in the large $D$ limit) $C = 1/(2\cdot \sqrt{e}) \approx 0.3033$ on high-girth graphs and gave numerical evidence that there is a one-local classical algorithm (building on \citet{shearer1992note,hirvonen2017large}) that dominates $\qaoa_1$ on all high-girth regular graphs. \citet{Marwaha2021localclassicalmax} gave numerical evidence that $\qaoa_2$ achieves $C < 0.41$ on high-girth graphs of sufficiently large $D$ and is dominated by a two-local classical algorithm (an extension of \citet{hastings2019classical}) on all high-girth regular graphs.

Much of the other work on understanding the QAOA focused on its \emph{worst-case approximation ratio}. 
For example, \citet{bravyi2019obstacles} constructed $D$-regular $n$-vertex bipartite graphs (where $1$ is the true maximum cut value) on which for every $p=o(\log_D n)$, $\qaoa_p$ achieves a cut of value at most $5/6 + O(1/\sqrt{D})$. \citet{farhi2020quantumWorstCase} showed that there are such bipartite graphs where $\qaoa_{o(\log_D n)}$ achieves a cut of value at most $1/2 + O(1/\sqrt{D})$. (Indeed, random bipartite graphs achieve this, since  $1-o(1)$ fraction of their local neighborhoods are tree-like; see also \citet{farhi2020quantumTypicalCase}.) 

However, as mentioned above, there are good reasons to believe that neither $\qaoa_p$ nor any efficient quantum algorithms (even ones that require scalable fault-tolerant quantum computers) can beat the best classical approximation ratio.
Thus, our focus is on \emph{per instance} comparisons of quantum and classical algorithms.  In other words, we ask whether in some particular family $\mathcal{I}$ of instances, there \emph{exists} $I \in \mathcal{I}$ on which local quantum algorithms such as the QAOA beat the results achievable by polynomial-time classical algorithms. \citet{bravyi2019obstacles} study the reverse question, and show that there exists some family $\mathcal{I'}$ of instances on which \citet{goemans1995improved}'s classical algorithm dominates $\qaoa_p$ for every constant $p$. 
Our theoretical results are for  graphs with high girth, though our experiments extend to graphs with short cycles, and we hope that future work will go beyond this limitation.

To our knowledge, all prior work on limitations of local maximum-cut algorithms (classical or quantum) used the method of \emph{indistinguishability}. That is, to show that an $r$-local algorithm $A$  outputs a cut of value at most $v$ on a graph $G$, one demonstrates a graph $G'$ that is locally indistinguishable from $G$ (in the sense that all or $1-o(1)$ fraction of $r$-local neighborhoods are isomorphic), and on which the true cut value is at most $v$. However, if random $D$-regular graphs minimize the maximum cut among all $D$-regular graphs with tree-like neighborhoods (which seems plausible), then this method cannot be used to rule out the possibility  that local algorithms can find cuts of value at least  $1/2 + P_*/\sqrt{D} -o(1/\sqrt{D})$ in high-girth graphs.
For maximum cut on random hypergraphs, \citet{Chen_2019} use the overlap gap property to prove that local algorithms are suboptimal; however, the overlap gap property is not expected to hold for the maximum cut problem on graphs, which is similar to the Sherrington-Kirkpatrick model in which it does not hold \citep{dembo2017extremal}.

Many works studied the complexity of sampling from the probability distribution induced by shallow circuits, much of it motivated by so called ``quantum advantage'' (also known as ``quantum supremacy'') proposals
\shortcites{terhal2004adptive,bremner2011classical,aaronson2011computational,aaronson2016complexity,bremner2016average,bermejo2017architectures,bravyi2018quantum,arute2019quantum,bouland2019complexity,farhi2019quantum,zhou2020limits,bouland2021noise}
\citep{terhal2004adptive,bremner2011classical,aaronson2011computational,aaronson2016complexity,bremner2016average,bermejo2017architectures,bravyi2018quantum,arute2019quantum,bouland2019complexity,farhi2019quantum,zhou2020limits,bouland2021noise}.

\citet{lightcone} and \cite{pan2021simulating} gave partial spoofing algorithms for the cross-entropy metric used in  \citet{arute2019quantum}'s quantum-advantage experiment; despite this, the assumption that sampling from the probability distribution is difficult is still well supported.
In particular, results for ``quantum sampling advantage'' of constant-depth quantum circuits include the following.
\citet{terhal2004adptive} proved that under widely believed complexity assumptions, there are depth  $4$ quantum circuits whose probability distribution cannot sampled precisely by a classical polynomial-time algorithm. 
\citet{markov2008simulating} gave a classical polynomial-time algorithm to simulate quantum circuits with logarithmic tree-width, but also showed that there exists a depth $4$ quantum circuit with linear tree width.  \citet{napp2019efficient} gave a classical algorithm for simulating random two dimensional circuits of some fixed constant depth, but also gave evidence that at some constant depth $d_0$, their algorithm undergoes a computational phase transition and becomes inefficient.
Perhaps most relevant to this work is the paper of \citet{farhi2019quantum} that gave evidence that even computing the probability distribution of $\qaoa_p$ for $p=1$ could be hard for classical algorithms.

\subsection{Discussion: NISQ optimization advantage}

\begin{figure}[htbp]
    \centering
    \includegraphics[width=5in]{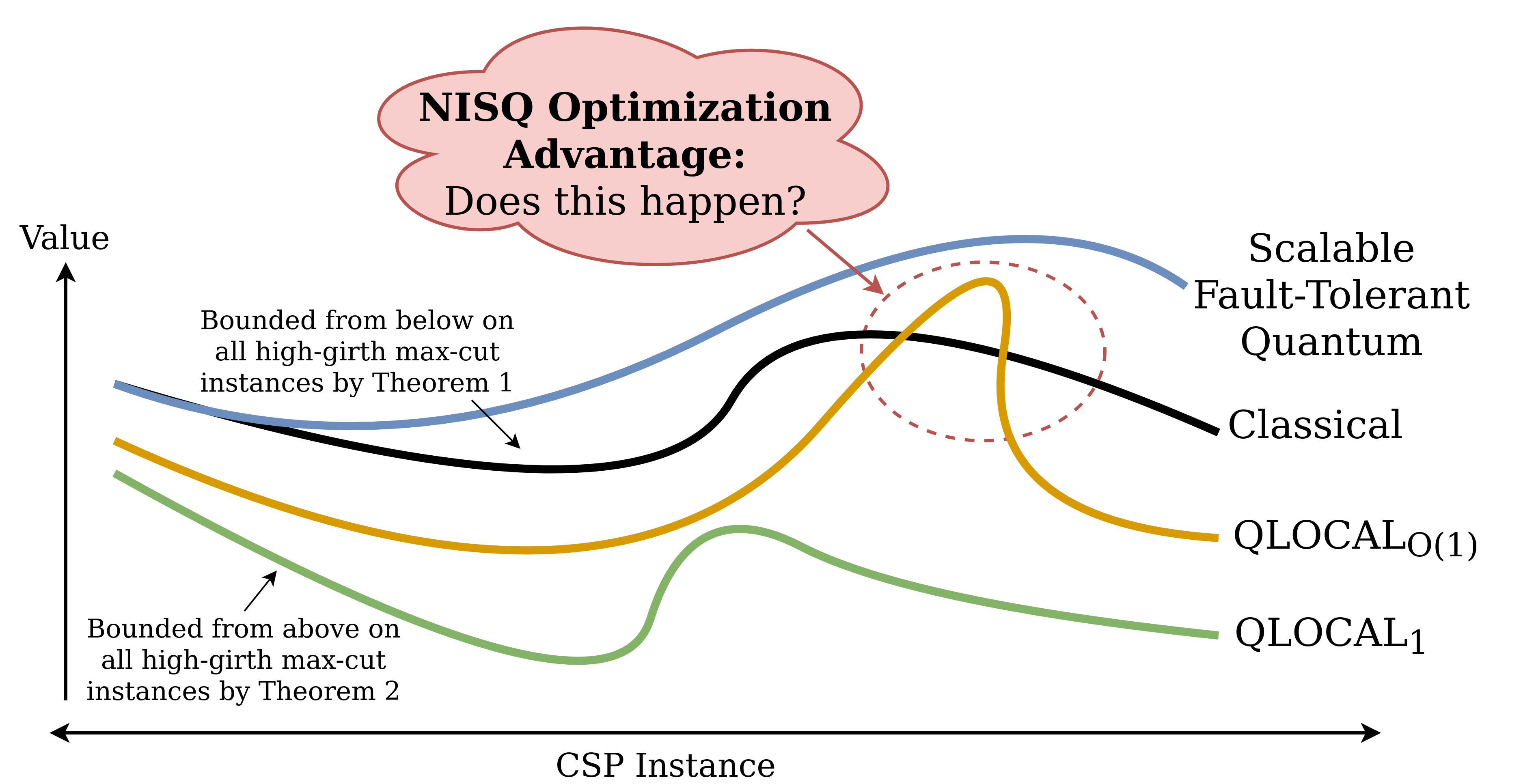}
    \caption{\footnotesize  Cartoon of performance of polynomial-time quantum, classical, and local algorithms on some constraint-satisfaction problem with the ``X axis'' consisting of all possible instances for the problem. Since quantum polynomial-time algorithms can simulate both classical polynomial time and constant-locality quantum algorithms, the corresponding curve (marked in blue)  is always equal or higher than all other curves, and is widely believed to be strictly above them for some instances (see Footnote~\ref{fn:factoring}). Theorem~\ref{thm:limit-informal} bounds the $\mathsf{QLOCAL}_1$ curve from above on high-girth maximum-cut instances, and Theorem~\ref{thm:algorithm-informal} bounds the classical curve from below on the same instances; we conjecture the latter bound can be improved to be strictly above $\mathsf{QLOCAL}_1$. \citet{bravyi2019obstacles,farhi2020quantumWorstCase} showed that there exist some instances on which local algorithms such as the QAOA perform strictly worse than classical polynomial-time algorithms. The question of ``NISQ optimization advantage'' can be phrased as asking whether there are any instances on which the $\mathsf{QLOCAL}_{O(1)}$ curve is above the classical curve.}
    \label{fig:nisq-advantage}
\end{figure}

We now discuss the relevance of our results to studying noisy intermediate-scale quantum devices (NISQ) devices, and the broader question of whether such devices can achieve a super-polynomial computational advantage over classical algorithms for optimization problems. This section contains no formal definitions or results used later on, and so can be safely skipped by readers interested in our technical results. See also Figure~\ref{fig:nisq-advantage} for a visual summary of this discussion. 

There is no agreed-upon formal definition of NISQ devices, but some characteristics of such devices include:

\begin{itemize}
    \item Computation happens across a fixed topology or graph $G_A$ (where $A$ stands for \emph{architecture}).
    
    \item Every gate involves a small number of qubits that are nearby in the topology.\footnote{Some NISQ hardware platforms can encode non-local ``hard constraints'', as in Section VII of \citet{farhi2014quantum}  (e.g., \citet{pichler2019quantum}). These platforms restrict the Hilbert space to feasible output states, and so can encode non-local unitaries. We do not consider such constraints in this work.}

    \item Every gate involves a constant amount of noise.
\end{itemize}

Given the above, as long as the noise in not low enough to allow for error correction \citep{aharonov2008fault,knill1998resilient,kitaev2003fault}, to ensure that most of the output qubits have more signal than noise, we need the number of operations (i.e., depth of computation) to be bounded by some constant depending on the noise level.\footnote{Here, we model NISQ devices as having a fixed amount of noise and a system that can scale to an arbitrary size. In current devices, the system size is relatively small compared to the noise level.} If an output qubit is computed using a small number of gates, then its ``light cone'' will only involve nearby vertices.
In general, even for optimization problems on graphs, the topology $G_A$ of the device's architecture need not be the same as the input graph $G$.
However, natural optimization algorithms such as QAOA perform best when the two match as closely as possible \citep{Harrigan2021}. So, since our focus here is on the \emph{limitations} of NISQ devices, it makes sense to consider the ``best case scenario'' where $G=G_A$.

In the above setting, NISQ algorithms for graph problems become the quantum version of the well known \textsf{LOCAL} model studied in distributed computing \citep{linial1992locality}.
The question of whether NISQ devices can obtain a super-polynomial advantage over classical algorithms in graph optimization problems is then formalized as follows.
We define a (hypergraph or graph) optimization problem $\varphi$ as a map that given the input (potentially labeled) graph $G=(V,E)$ and some assignment $x \in D^V$, outputs a number $\varphi_G(x) \in [0,1]$, where the goal is to find, given the instance $G$, the assignment $x$ that maximizes $\varphi_G(x)$.
We now say that $\varphi$ exhibits a \emph{NISQ optimization advantage} if there exists an $O(1)$-local quantum algorithm $A$ and $\epsilon>0$ such that for every classical polynomial-time algorithm $B$, there exists some instance $G$ such that $\varphi_G(A(G)) > \varphi_G(B(G))+\epsilon$.\footnote{If we restrict our attention to uniform algorithms, we can use results such as Levin's universal search algorithm to argue that in such a case there would be an instance on which $A$ dominates all polynomial-time classical algorithms \citep{levin1973universal}. Otherwise we can modify the condition to ask for a set $\mathcal{I}$ of instances on which the average performance of $A$ is $\epsilon$ higher than the average performance of all polynomial-time classical algorithms.}
This is in some sense a ``best case complexity'' analysis of NISQ, since it does not require that the instances on which the device beats all classical algorithms are useful in any way. Nevertheless, at the moment we are not even able to rule this out. Theorem~\ref{thm:limit-informal} strongly indicates that at least for high-girth maximum-cut instances and one-local quantum algorithms, no such advantage exists.\footnote{We say ``strongly indicates'' since at the moment our classical algorithm gives a constant $C = 2/\pi \approx 0.6366$ that is smaller than the constant $1/\sqrt{2} \approx 0.7071$  ruled out by Theorem~\ref{thm:limit-informal}. However, as mentioned above, we conjecture that the classical algorithm can be improved to give the constant $P_* \approx 0.7632$ which is above what is achievable by one-local quantum algorithm.}
See Figure~\ref{fig:nisq-advantage} for an illustration of the question of ``NISQ optimization advantage'' for general CSPs.

\section{Classical and quantum one-local algorithms are suboptimal}

In this section we prove Theorem~\ref{thm:limit-informal}, showing that every (quantum or classical) one-local algorithm achieves a cut value of at most $\tfrac{1}{2} + \tfrac{1}{\sqrt{2D}}$ on high-girth regular graphs. 
To state the result formally, we define the notion of \emph{local distributions}.
Specifically we will show that:

\begin{itemize}
    \item Every one-local algorithm for maximum cut induces a one-local distribution. (This uses standard arguments, and is deferred to Appendix~\ref{app:local}.)
    
    \item For every centered\footnote{The assumption of ``centeredness'' is discussed below. It is both minimal, in the sense that it is satisfied by all natural randomized local algorithms, including QAOA, as well as necessary to rule out pathological examples.} one-local distribution over cuts in a high-girth $D$-regular graph, the expected cut value is at most $\tfrac{1}{2} + \tfrac{1}{\sqrt{2D}}$. (This is the technical heart of the proof.)
\end{itemize}

\subsection{Local distributions}
\label{sec:local}
A randomized algorithm $A$ for maximum cut takes as input a graph $G=(V,E)$, and produces a probability distribution $X$ over $\{ \pm 1 \}^V$. The \emph{value} of the cut is
$$
\Pr_{(i,j) \in E}[ X_i \neq X_j] = \tfrac{1}{2} - \tfrac{1}{2}\E_{(i,j)\in E, X} [X_iX_j] 
$$
The central notion we will use in this paper is that of \emph{local distribution}. 
We use the following notation: If $G = (V,E)$ is a graph, $S \subseteq V$ is a set, and $r\in \mathbb{N}$, then we let $B_r(S)$ denote the ball of radius $r$ around $S$, namely the set of vertices that are of distance $\leq r$ to $S$. (In particular, $S \subseteq B_r(S)$.)  If $X$ is a  distribution over $\{ \pm 1 \}^V$ and $S \subseteq V$, then $X_S$ is the marginal distribution over the set $\{ \pm 1 \}^S$.

\begin{definition}[Local distributions] \label{def:local-dist} 
Let $G=(V,E)$ be a graph and $X$ be a distribution over $\{ \pm 1 \}^V$. For every $r \in \mathbb{N}$, We say that $X$ is \emph{$r$-local} if for every sets $A,B \subseteq V$, if $B_r(A) \cap B_r(B) = \emptyset$ then $X_A$ is independent from $X_B$. We say that the distribution is \emph{centered} if $\E[X] = 0^V$. 
\end{definition}

As we show in Appendix~\ref{app:local}, every $r$-local quantum or classical algorithm for maximum cut induces an $r$-local distribution on its output. This is not an equivalence between local algorithms and distributions: the locality of algorithms induces more conditions on the output distribution than Definition~\ref{def:local-dist}, and in particular the outputs of local classical algorithms are more restricted than the outputs of local quantum algorithms.  However, since our focus is obtaining negative results, it suffices to restrict attention to local distributions.

\paragraph{Centered distributions.} Because of the symmetry in the problem itself (where $-X$ is a cut of the same value as $+X$), all natural randomized algorithms for maximum cut (including QAOA, ALR, and \cite{goemans1995improved}'s algorithm) induce a centered distribution, and we will assume this condition in what follows. Because we aim to give \emph{pointwise} lower bounds, such  an assumption is also necessary to rule out the trivial $0$-local algorithm that always outputs a fixed cut $x_0$ that is the optimal cut for some particular instance $G_0$. 

\medskip Our  negative result is the following:

\begin{thm}[Formal version of Theorem~\ref{thm:limit-informal}] \label{thm:limit-formal} Let $G=G(V,E)$ be an $D$-regular graph of girth at least $6$. Then, for every centered one-local centered distribution $X$ over $G$, 
\[
\E_{(i,j)\sim E,X} \left[ X_iX_j \right] \geq -\tfrac{2C}{\sqrt{D}} \;,
\]
where $C=1/\sqrt{2} \approx 0.7071$.
\end{thm}

\paragraph{Symmetry.} Output distributions arising from natural local algorithms satisfy a stronger notion of symmetry, which is that the identities of vertices and their neighbors do not affect the marginal distributions. This means that for every set $A = \{ a_1,\ldots, a_\ell \}$ of vertices, if $\psi:B_r(A) \rightarrow V$ is an isomorphism of the graph (one-to-one function such that $(u,v)$ is an edge iff $(\psi(u),\psi(v))$ is an edge), the marginal distributions $X_{a_1,\ldots,a_\ell}$ and $X_{\psi(a_1),\ldots,\psi(a_\ell)}$ are identical. We do not use the assumption of symmetry in this work, but it may be useful for proving stronger negative results.

\subsection{Proof of Theorem~\ref{thm:limit-formal}}

Consider a $D$-regular $n$-vertex graph $G$ with no cycles shorter than $6$ and some centered one-local probability distribution $X$ over $\{ \pm 1 \}^n$. The probability distribution satisfies that for every pairs of vertices $u$ and $v$ that are of distance at least three, $E[ X_uX_v] = 0$ and that for every edge $u,v$, $E[X_uX_v] = -2C/\sqrt{D}$. This implies a cut of value  $1/2 + C/\sqrt{D}$. We want to upper bound $C$. We define the following notation:
\begin{itemize}
    \item For vertex $u$ and $\sigma \in \{ \pm 1 \}$, $\mu_{u,\sigma} := E_{v \sim u} [ X_uX_v | X_u = \sigma ]$.  Note that $E_{u,\sigma} [\mu_{u,\sigma}] = -2C/\sqrt{D}$ where expectation is taken over a random vertex $u$ and random sign $\sigma$ (since $X$ is centered, the marginal for every vertex is always uniform).
    \item Define $\mu_{u,\sigma}^{(2)}$ as the expectation of correlation for a 2 step random walk where $u$ is the \emph{second} vertex in the walk, conditioned on $X_u = \sigma$. That is, $\mu_{u,\sigma}^{(2)} = E_{a\sim u \sim b}[X_a X_b|X_u = \sigma]$.
    \item Define $\mu_{u,\sigma}^{(3)}$ as the expectation of correlation for a 3 step random walk with $u$ as the first vertex, conditioned again on $X_u = \sigma$. That is $\mu_{u,\sigma}^{(3)} := E_{u \sim b \sim c \sim d} [ X_uX_d | X_u = \sigma ]$. 
    \item Define $\mu_{u,\sigma}^{(4)}$ similarly for a 4 step random walk  but where $u$ is again the \emph{second} vertex in the walk. That is, $\mu_{u,\sigma}^{(4)} = E_{a \sim u \sim b \sim c \sim d}[ X_aX_d | X_u = \sigma ]$.
\end{itemize}
We now make the following claims:

\begin{claim}
$\mu_{u,\sigma}^{(3)} = \tfrac{2-1/D}{D}\mu_{u,\sigma}$.
\end{claim}
\begin{pf}
Because the girth is at least 6, a 3 step random walk locally looks like a walk on a tree. It can either go to distance 1 (if the second or third edges are back edges, which happens with probability $1/D$ and $(1-1/D)/D$ respectively) or to distance 3. If it goes to distance 1 then we get a contribution of $\mu_{u,\sigma}$. By one-locality and centeredness, if $u$ and $d$ are of distance at least $3$, then the marginals of $u$ and $v$ are uniform and independent, so conditioned on $X_u=\sigma$, $\E[X_uX_d] = \sigma \E[X_d] = 0$ . 
\end{pf}

\begin{claim}
$\mu_{u,\sigma}^{(4)} = \mu_{u,\sigma} \cdot \mu_{u,\sigma}^{(3)}$.
\end{claim}
\begin{pf}
A 4 step walk where $u$ is the second step can be thought of as taking independently 1 step from $u$ and $3$ steps from $u$.
In expectation the endpoint of one step will be $\mu_{u,\sigma} \cdot \sigma$  and the endpoint of the $3$ step walk will be $\mu_{u,\sigma}^{(3)} \cdot \sigma$. Since they are independent, expectation of the product is the product of expectations and since $\sigma^2=1$ we get the result.
\end{pf}
\begin{claim}
$\mu_{u,\sigma}^{(4)} = \tfrac{2-1/D}{D} \mu_{u,\sigma}^2$.
\end{claim}
\begin{pf}
This is implied by Claims 1 and 2.
\end{pf}
\begin{claim}
$\mu_{u,\sigma}^{(2)} = \mu_{u,\sigma}^2$.
\end{claim}
\begin{pf}
Similarly to Claim 2, a 2 step walk where $u$ is the second step can be thought of as taking two independent 1 step paths from $u$. In expectation, the endpoint of each step is $\mu_{u,\sigma} \cdot \sigma$. Since the endpoints are independent, the expectation of the product is the product of the expectations and since $\sigma^2 = 1$ we get the result.
\end{pf}

We now average over all the choices of $u$ and $\sigma$.
\begin{claim}
For $k\in\{1,2\}$, $E_{u,\sigma}[\mu_{u,\sigma}^{(2k)}] = E_x[x^\top A^{2k} x]$ where $x = X/\sqrt{n}$ and $A$ is $1/D$ times the adjacency matrix of the graph.
\end{claim}
\begin{pf}
Since $A$ is the random-walk matrix, for every $x\in \{ \pm 1\}^n$, the right-hand side equals the sum over all $i,j$ of the probability that $j$ is reached from $i$ via a random $2k$ step walk times $x_ix_j/n$, and hence equals the expectation of $X_iX_j$ where $i$ is a random vertex and $j$ is obtained by taking a $2k$ step random walk from $i$. 

The left-hand side corresponds to the expectation of the following quantity:
\begin{itemize}
    \item We pick vertex $u$ at random and $\sigma \in \{ \pm 1 \}$.
    \item We pick a neighbor $a \sim u$ at random, and a $(2k-1)$-path from $u$ ($u \sim ... \sim d$) at random.
    \item We output $X_aX_d | X_u = \sigma$.
\end{itemize}
Since the marginal $X_u$ is uniform over $\{ \pm 1 \}$ this is the same as picking $u$ at random and let $\sigma = X_u$, and then repeating the same process, in which case we can drop the conditioning. Since the graph is regular, the induced distribution on $a,d$ is identical to that of endpoints of a random $2k$ step path. So, the result holds.
\end{pf}
\begin{claim}
$E_{u,\sigma}[\mu_{u,\sigma}^{(4)}] \geq \left( E_{u,\sigma} [\mu_{u,\sigma}^2] \right)^2$.
\end{claim}
\begin{pf}
Let $(v_1,\ldots,v_n)$ be the normalized eigenvectors of $A$. Then every unit vector $x$ can be written as $x = \sum_{i=1}^n \alpha_i v_i$ where $\sum_{i=1}^n \alpha_i^2 = 1$. Hence $x^\top A^4 x = \sum_{i=1}^n \alpha_i^2 \lambda_i^4 = \E[ \lambda^4]$ where $\lambda$ is the random variable where $\Pr[ \lambda = \lambda_i ] = \alpha_i^2$. By convexity $\E[ \lambda^4 ] \geq \E[ \lambda^2 ]^2$ and so for every unit vector $x$, $x^\top A^4 x \geq \left(x^\top A^2 x\right)^2$. 

Hence 
$$
E_{u,\sigma} [\mu_{u,\sigma}^{(4)}]
= E_x[ x^\top A^4 x]
\geq  E_x [\left( x^\top A^2 x \right)^2] \geq \left( E_x[ x^\top A^2 x] \right)^2 = \left(E_{u,\sigma}[\mu_{u,\sigma}^{(2)}] \right)^2 = \left(E_{u,\sigma}[\mu_{u,\sigma}^2] \right)^2
$$ 
with the second inequality following from Cauchy-Schwarz and the last equality from Claim 4.
\end{pf}
Combining Claims 3 and 6 we get that $\left( E_{u,\sigma} [\mu_{u,\sigma}^2] \right)^2 \leq \tfrac{2-1/D}{D}\left( E_{u,\sigma} [\mu_{u,\sigma}^2] \right)$ and so $E_{u,\sigma} [\mu_{u,\sigma}^2] \leq \tfrac{2-1/D}{D}$.
Using Cauchy-Schwarz, this implies $\bigl| E_{u,\sigma} [\mu_{u,\sigma}] \bigr|  \leq \sqrt{\tfrac{2-1/D}{D}}$
which means $C \leq 0.5\sqrt{2-1/D} < 1/\sqrt{2}$.

\section{A classical algorithm for maximum cut on high-girth graphs}

In this section we prove Theorem~\ref{thm:algorithm-informal}. That is, we show that there is a polynomial time algorithm that, given a high-girth $D$-regular graph, finds a cut of at least $\tfrac{1}{2} + \tfrac{2}{\pi \sqrt{D}} - o(1)$  (where the $o(1)$ term tends to zero with $D$, the girth, and the running time of the algorithm). 
This is an algorithmic version of the bound of \citet{lyons2017factors}, who proved that there exists a cut of this magnitude for every high-girth graph.
The algorithm is in fact a $k$-local classical algorithm, with the value of the cut improving with $k$ and the girth.

We now restate Theorem~\ref{thm:algorithm-informal} more formally  and prove it.

\begin{thm}[Formal version of Theorem~\ref{thm:algorithm-informal}] \label{thm:algorithm} 
For every $k$, there is a $k$-local algorithm $A$ such that for all $D$-regular $n$-vertex graphs $G$ with girth $g > 2k+1$, $A$ outputs a cut $x \in \{\pm 1\}^n$ cutting $\cos^{-1}(-2\sqrt{D-1}/D)/\pi - O(1/\sqrt{k}) > 1/2 + 2/(\pi\sqrt{D}) - O(1/\sqrt{k})$ fraction of edges.
\end{thm}

\begin{pf}
Our algorithm is as follows:

\begin{enumerate}
    \item Assign every vertex $w$ a value $Y_w \sim \mathcal{N}(0,1)$.
    \item For every vertex $u$,  let $X_u = \sgn\Big(\sum_{w; d(w,u)\le k} (-1)^{d(w,u)} (D-1)^{-0.5d(w,u)}Y_w\Big)$
where $d(w,u)$ is the graph distance from $u$ to $w$.
    \item Output the vector $X$.
\end{enumerate}

\begin{figure}[htbp]
    \centering
    \includegraphics[width=3in]{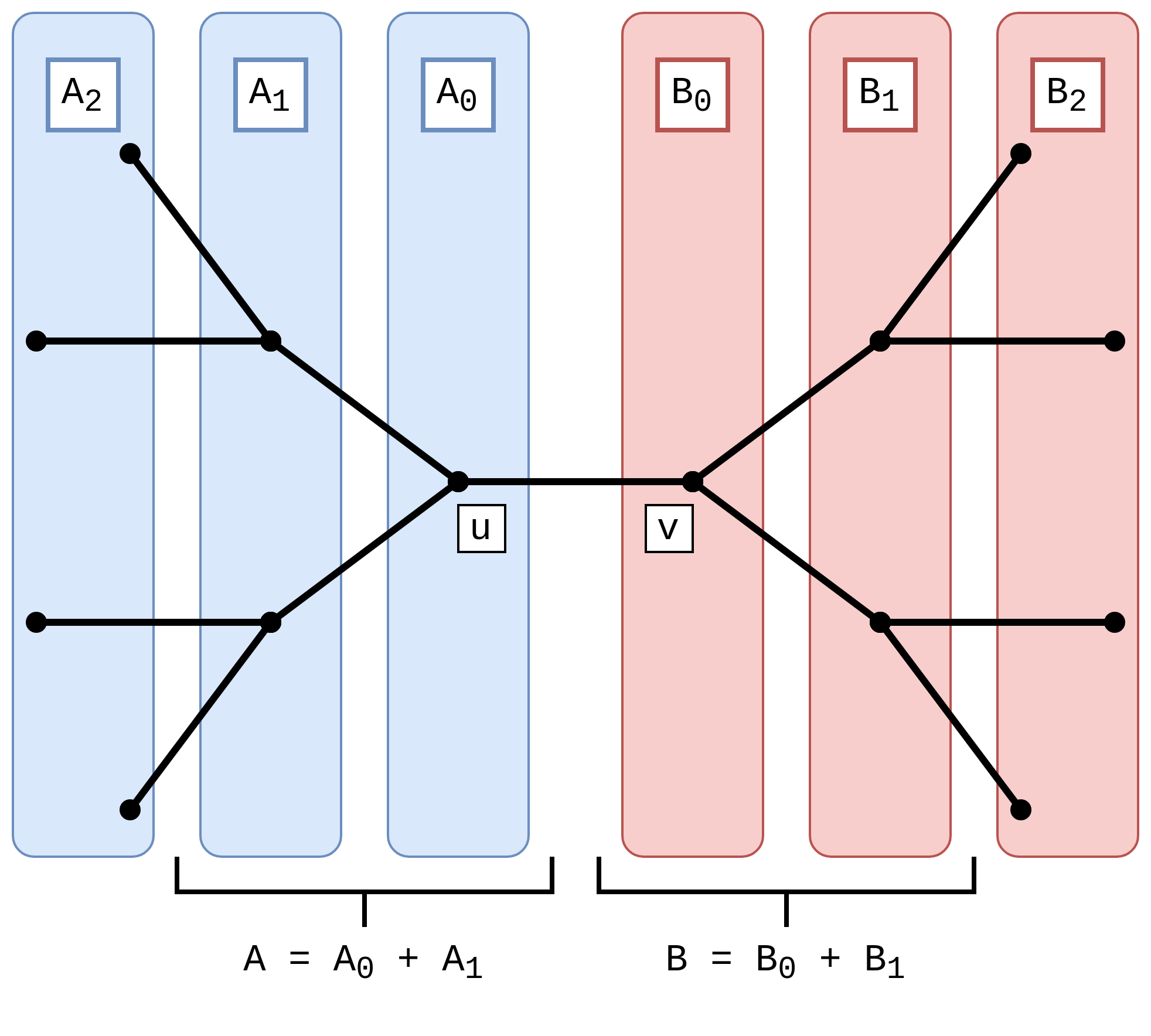}
    \caption{\footnotesize Analysis of the algorithm when $k=2$ and $D=3$. $A_\ell$ involves the nodes closer to $u$, and $B_\ell$ involves the nodes closer to $v$. Since all nodes are at most distance $2k+1$ from each other (which is smaller than the graph's girth), the shortest paths from $E_{uv}$ to other vertices form a tree.}
    \label{fig:klocal}
\end{figure}

\paragraph{Analysis.} Consider the radius $k$ neighborhoods around vertices $u$ and $v$ for some edge $E_{uv}$. Since the graph's girth is more than $2k+1$, and all nodes are within distance $k+1$ of $u$, the subgraph locally looks like two depth-$k$ trees rooted at $u$ and $v$.\footnote{There can be a $2k+2$ cycle that connects the leaves of the $u$ tree to the leaves of the $v$ tree, but such a cycle does not create any ``shortcut'' that is not encoded in the trees, and so we can ignore it.} Consider these trees, then define
\begin{align*}
    A_\ell &= \sum_{w; d(u,w) = \ell<d(v,w)} (-1)^\ell (D-1)^{-0.5\ell} Y_w
    & 
    B_\ell &= \sum_{w; d(v,w) = \ell<d(u,w)} (-1)^\ell (D-1)^{-0.5\ell} Y_w    
\end{align*}
Here, $A_\ell$ is summing up nodes closer to $u$, and $B_\ell$ is summing up nodes closer to $v$. See Figure \ref{fig:klocal} for an illustration. Since $A_\ell$ is a sum of $(D-1)^\ell$ independent normal variables with variance $(D-1)^{-\ell}$, $A_\ell$ is a standard normal variable. Same for $B_\ell$. Let's also define the following:
\begin{align*}
    A &= \sum_{\ell=0}^{k-1} A_\ell
    & 
    B &= \sum_{\ell=0}^{k-1} B_\ell
    \\
    U &= A_k + A - B/\sqrt{D-1}
    & 
    V &= B_k + B - A/\sqrt{D-1}
\end{align*}
Then $X_u = \sgn(U)$ and $X_v = \sgn(V)$. Since $A, B, A_k, B_k$ are all individual Gaussians, we claim
$$
Pr[\sgn(U) \ne \sgn(V)] \ge \frac{1}{2} + \frac{2}{\pi}\tan^{-1}(\frac{1}{\sqrt{D-1}}) - O(\frac{1}{\sqrt{k}})
$$
Since $|A_k| \le |A|/\sqrt{k}$ and $|B_k| \le |B|/\sqrt{k}$, ignoring the $A_k$ and $B_k$ only affects the probabilities by $O(1/\sqrt{k})$. For example, $\sgn(U) = \sgn(U - A_k)$ except possibly when $|U| \le O(\sigma_{A_k})$; the sign can differ only when the magnitude of $U$ is within a few standard deviations of $A_k$. This happens with probability $\erf(O(1/\sqrt{k})) = O(1/\sqrt{k})$. So ignoring the $A_k$ and $B_k$ terms will lose at most $O(1/\sqrt{k})$ fraction of edges.

Consider the probability $S$ that $\sgn(A - B/\sqrt{D-1}) \ne \sgn(B - A/\sqrt{D-1})$ for i.i.d normal variables $A, B \sim \mathcal{N}(0,k)$. The variance should not affect the sign, so this should match the probability that $\sgn(P - Q/\sqrt{D-1}) \ne \sgn(Q - P/\sqrt{D-1})$ for standard normals $P, Q \sim \mathcal{N}(0,1)$. For $p \sim P$ and $q \sim Q$, this is false when $p > q/\sqrt{D-1} > p/(D-1)$ or $p < q/\sqrt{D-1} < p/(D-1)$. This requires $p$ and $q$ to have the same sign, so $1/\sqrt{D-1} < p/q < \sqrt{D-1}$ for $D>1$. The chance of this happening is
\begin{align*}
    1-S &= \frac{2}{\sqrt{2\pi}} \int_{0}^\infty dx\ e^{-x^2/2} \frac{1}{\sqrt{2\pi}} \int_{x/\sqrt{D-1}}^{x\sqrt{D-1}}dy\  e^{-y^2/2}
    \\
    &= \frac{1}{\sqrt{2\pi}}\int_{0}^\infty dx\ e^{-x^2/2} 
    \Big( \erf(\frac{x}{\sqrt{2}} \sqrt{D-1})
    - \erf(\frac{x}{\sqrt{2}} \frac{1}{\sqrt{D-1}})
    \Big)
    \\
    &= \frac{1}{\sqrt{2\pi}} \sqrt{\frac{2}{\pi}}\Big(\tan^{-1}(\sqrt{D-1}) - \tan^{-1}(\frac{1}{\sqrt{D-1}}) \Big)
    \\
    &= \frac{1}{\pi} \Big( \frac{\pi}{2} - 2\tan^{-1}(\frac{1}{\sqrt{D-1}}) \Big)
\end{align*}
So the probability $S$ is $\frac{1}{2} + \frac{2}{\pi} \tan^{-1}(\frac{1}{\sqrt{D-1}})$, which gives, for example, $S \ge \frac{1}{2} + \frac{2}{\pi}\frac{1}{\sqrt{D}}$. Some algebraic manipulation shows that $S = \cos^{-1}(-2\sqrt{D-1}/D)$, matching the result in \citet{lyons2017factors}.
\end{pf}

\begin{figure}[p]
    \centering
    \hspace*{-0.5in}
    \includegraphics[width=7.5in]{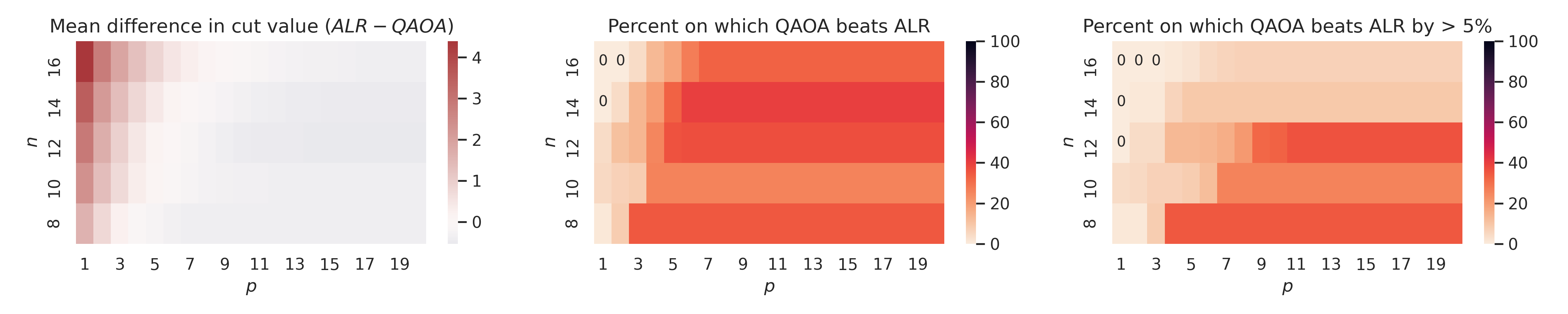}
    \caption{\footnotesize Comparing the ALR algorithm and $\qaoa_p$ (for $p=1\ldots 20$) on random 3-regular graphs of varying sizes ($n=8,10,12,14,16$). Left: The average (across instances and QAOA measurements) difference between the ALR and QAOA cut values. Middle: The fraction (in percent) of instances on which QAOA achieves a better value than ALR. Right: Fraction of instances on which QAOA achieves a value at least $5\%$ better than ALR. In both the middle and right panel, a fraction of $0$ means that ALR outperformed (respectively nearly outperformed) QAOA on \emph{every} instance for those values of $n$ and $p$. We see that even for $p>1$, if $n$ is sufficiently large relative to $p$ then ALR starts to dominate $\qaoa_p$. All QAOA simulations are taken from \citet{Zhou_2020}. As $p$ grows large enough compared to $n$, QAOA eventually finds the true maximum cut.}
    \label{fig:random}
\end{figure}

\begin{figure}[p]
    \centering
    \includegraphics[width=5in]{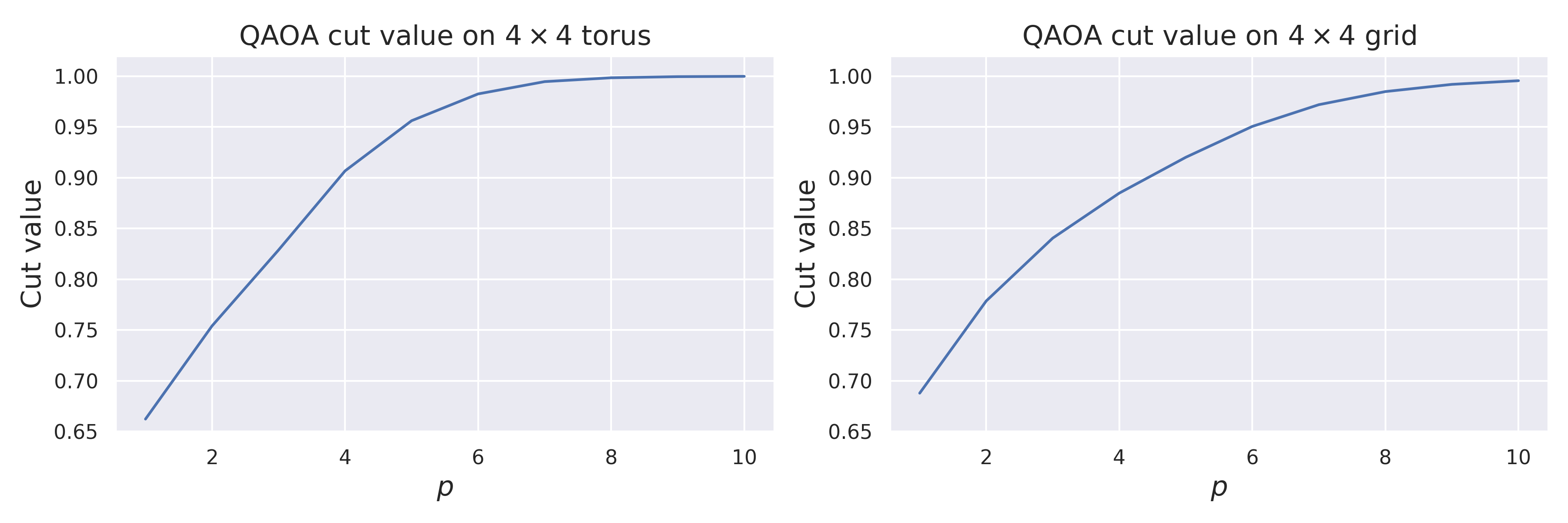}
    \caption{\footnotesize Performance of $\qaoa_p$ on the $4\times 4$ grid and torus graphs. These are bipartite graphs on which the ALR algorithm finds the perfect cut of value $1$. We see that for $p \ll n$, $\qaoa_p$ fails to find the maximum cut despite the existence of small cycles. Code taken from \citet{Zhou_2020}.}
    \label{fig:grid}
\end{figure}

\begin{figure}[p]
    \centering
    \includegraphics[width=3in]{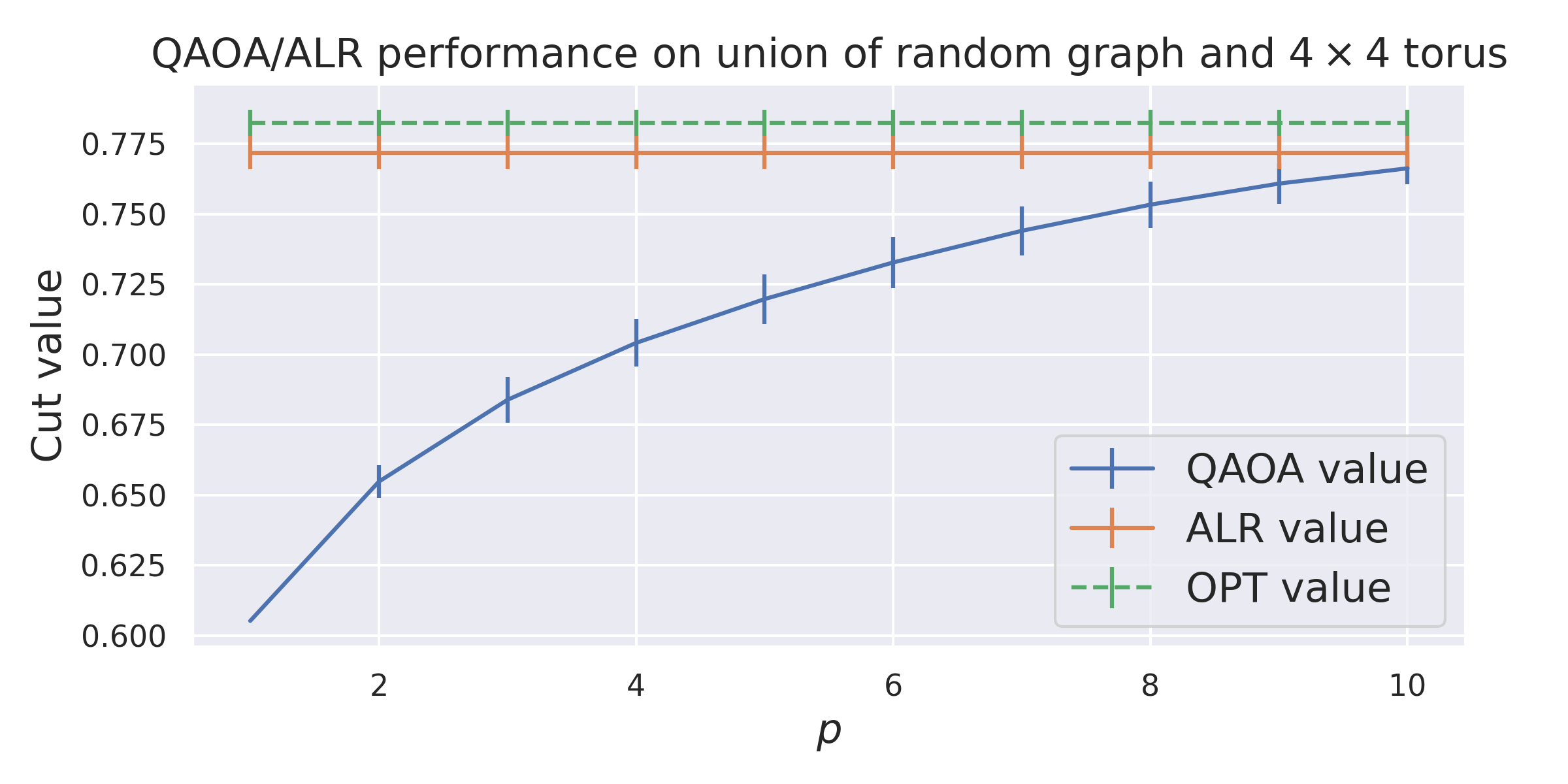}
    \caption{\footnotesize Performance of ALR and QAOA on the union of a random $16$ vertex graph and the $4\times 4$ torus. Due to computational constraints, we only ran this simulation on two graphs, and so the error bars correspond to the difference between the highest (resp. lowest) and mean values in the two graphs.}
    \label{fig:union}
\end{figure}

\paragraph{Remark: Relationship to the ALR algorithm.}
At large $k$, this algorithm has the same performance as the algorithm by \citet*{Aizenman1987} for the Sherrington-Kirkpatrick (SK) model.  For optimizing the SK model the goal is, given an $n\times n$ matrix $A$ (sampled from the Gaussian Orthogonal Ensemble\footnote{The Gaussian Orthogonal Ensemble or GOE is the probability distribution on $n\times n$ matrices $B$ obtained by letting $B=(A+A^\top)/\sqrt{2}$ where $A$ is a matrix for which $A_{i,j}$ is independently sampled from a standard normal variable for every $i,j \in [n]$.}), to find a vector $x \in \{ \pm 1 \}^n$ that minimizes $x^\top A x$. The ALR algorithm is to let $x$ be the sign of the minimum eigenvector of $A$. The algorithm of Theorem~\ref{thm:algorithm} is similar (though not identical) to taking $x$ to be the sign of $(I-\tfrac{1}{\sqrt{D-1}}A)^k y$ where $A$ is the adjacency matrix of the graph and $y$ is a standard Gaussian vector, and hence can be thought of as a truncated version of the power-method computation of the minimum eigenvector.\footnote{The two algorithms are not identical since $A^k$ also accounts for walks  that include ``back edges''.}

Just like the ALR algorithm is not optimal for the SK model \citep{montanari2021optimization}, we suspect that the algorithm of Theorem~\ref{thm:algorithm} is not optimal either, and that there is a classical polynomial time algorithm that achieves a cut of value $1/2 + P_*/\sqrt{D} + o(1/\sqrt{D})$ for every $D$-regular graph of sufficiently high girth. 

\section{Beyond one-locality and high girth: some computational experiments} \label{sec:experiments}

Our classical algorithm is only analyzed for graphs with high girth, while our negative results are only established for one-local algorithms. In this section we present some experimental results that indicate that the results are likely to extend at least somewhat beyond these bounds. 
These results are also described in the Jupyter notebook \url{http://tiny.cc/QAOAvsALR}.
For starters, let us show that we cannot expect the ALR algorithm, nor the semidefinite program of \citet{goemans1995improved} (GW), to beat $\qaoa_{O(1)}$ on \emph{every} graph.

\begin{thm} \label{thm:QAOAbeatsClassical}
There exists some $\epsilon>0$ and $p\in\mathbb{N}$, and an infinite sequence of graphs $\{ G_m \}$  such that for every $m$, 
$val(\text{QAOA}_p(G_m)) > \max \{ val(ALR(G_m)), val(GW(G_m)) \} + \epsilon$.
\end{thm}
\begin{pf}[Proof Sketch] We sketch the proof for the GW program, though the idea is similar for the $ALR$ algorithm. It is known that there is a fixed graph $G_0$ of size $n_0$ on which the GW algorithm produces a cut that is some constant $\epsilon_0>0$ smaller than the optimum \citep{karloff1999good}.  
It is also known that in the limit of $p \rightarrow \infty$, $\qaoa_p$ achieves the optimum value for every input \cite[Eq.~(10)]{farhi2014quantum}. Hence there is some $p_0 = p_0(n_0)$ on which $\qaoa_{p_0}$ achieves a value that is $\epsilon_0>0$ larger than GW. By the nature of both algorithms, the (fractional) value of the cut they produce on a disjoint union of copies of $G_0$ will be the same as the value they produce on $G_0$, and so the family $\{ G_m \}$ will be of graphs that are composed of $m$ disjoint copies  of $G_0$.
\end{pf}

Note that Theorem~\ref{thm:QAOAbeatsClassical} does not preclude that there is a classical algorithm that can match or beat $\qaoa_p$ on every graph for every $p=O(1)$ (or even $p$'s that grow with $n$). However, it does show that neither the ALR nor the GW algorithm can do so.  Nevertheless, it is still interesting to find out the answer to the following questions: 

\begin{enumerate}
    \item Does the ALR algorithm beat $\qaoa_p$ for values of $p$ larger than $1$ on random regular graphs or high-girth graphs?
    
    \item Does the ALR algorithm beat $\qaoa_p$ on natural examples of graphs with small cycles, such as the grid?
\end{enumerate}

We describe our computational experiments in Figures~\ref{fig:random}, \ref{fig:grid}, \ref{fig:union}. We have taken some of the instances on which QAOA was simulated by \citet{Zhou_2020} (and which were generously shared with us by the authors) and compared the performance of ALR on the same instances ($100$ random unweighted 3-regular graphs for each of $N\in \{8,10,12,14,16\}$). These results suggest that as the size $n$ grows relative to the QAOA depth $p$, the relative performance of QAOA deteriorates. This offers more evidence for the intuition (also arising from \citet{farhi2020quantumTypicalCase,farhi2020quantumWorstCase}) that for $\qaoa_p$ to beat classical algorithms, a necessary condition is for $p$ to grow with $n$. 
This is in contrast to the SK model at infinite size \citep{qaoa_skmodel}, where QAOA at $p=11$ can surpass the ALR algorithm (though not the classucal algorithm of \citet{montanari2021optimization}).
Further investigation is needed to see if there exist maximum-cut instances on which constant-depth QAOA surpasses all polynomial-time classical algorithms.

To check the significance of the girth condition, we also compare the performance of ALR and QAOA on the grid and torus graphs which are arguably the prototypical graphs with small cycles.
For the grid and the even side length torus, these graphs are bipartite and their smallest eigenvector corresponds to this bipartition, so the ALR algorithm finds the optimal cut.\footnote{For odd-sized $n\times n$ torus, the ALR algorithm finds a cut of value $1-o(1)$ where $o(1)$ tends to zero with $n$.} Hence the question is the value of QAOA on these graphs. Figure~\ref{fig:grid} presents simulations of QAOA on these graphs, suggesting that as $p \ll n$, the value of the cut found by QAOA is bounded away from $1$. We used the code of \citet{Zhou_2020}, as well as their hyperparameter choices.
We also tried adding small cycles to the random graphs used in the simulations of Figure~\ref{fig:random} by superimposing a $16$-vertex random $3$-regular graph on the $4\times 4$ torus, and obtained similar results: see Figure~\ref{fig:union} for details.

\section{Conclusions}

In the near term, \emph{depth} and \emph{locality} are likely to be highly restricted resources for quantum computation. This work points at the possibility that such restrictions are at odds with obtaining quantum advantage for optimization problems, not just in the worst case but for every possible instance. However, our theoretical results are at the moment extremely limited. Improving the classical algorithm to achieve the optimal value of $1/2 + P_*/\sqrt{D}$  and extending the negative results to handle $k$-local algorithms for $k>1$ are the most immediate open questions. Other directions include generalizing beyond maximum cut, and finding natural classical algorithms to compete with QAOA and other local quantum algorithms that are not subject to the limitations of Theorem~\ref{thm:QAOAbeatsClassical} and could potentially \emph{pointwise dominate} all local quantum algorithms. One natural candidate for such an algorithm is the \emph{sum of squares} algorithm~\citep{Lasserre00,Parrilo00,barak2014sum}. Understanding the power of $\qaoa_p$ for non-constant but slowly growing values of $p$ (e.g. $p=O(\log n)$) is also an important open question. 

\section*{Acknowledgements}
We thank Beatrice Nash for collaborating on early stages of this project.
KM thanks Jeffrey M. Epstein and Matt Hastings for explaining details on Lieb-Robinson bounds.  Matt Hastings proposed a question similar to this one. Ruslan Shaydulin offered suggestions on simulating QAOA. We thank Madelyn Cain, Eddie Farhi, Pravesh Kothari, Sam Hopkins, and Leo Zhou for useful discussions. Special thanks to Leo Zhou for sharing with us the code and data from the paper \citep{Zhou_2020}.

\bibliography{research}

\appendix
\section{Local algorithms induce local distributions}
\label{app:local}
Here, we prove that every $r$-local quantum or classical algorithm for maximum cut induces a  $r$-local distribution on its output.

\paragraph{Classical algorithms.} A randomized $r$-local classical algorithm for maximum cut has the following form. For every vertex $v$ in the graph we choose some random variable $Y_v$ independently from some distribution $D$ over some domain $S$. Then, on input a graph $G$, the algorithm outputs $X \in \{ \pm 1\}^V$ defined as follows. For every $u \in V$, $X_u$ is obtained by some function $f(H_u)$ where $H_u$ is the labeled graph obtained by taking the radius $r$ ball around $u$ and labeling each vertex $v$ in this graph with $Y_v$. This corresponds to the randomized local model (sometimes denoted as   \textsf{RandLOCAL}) in distributed computing \citep{linial1992locality}. Such an algorithm also corresponds to a  \emph{factor of independent and identically distributed variables}, known as a FIID~\citep{correlationdecay,lyons2017factors,Chen_2019}.

\begin{thm}[Local classical algorithms induce local distributions]
Consider an $r$-local classical algorithm for maximum cut. Given any graph $G$, the algorithm will output a cut from an $r$-local distribution.
\end{thm}
\begin{pf}
Consider vertices $x, y$ where $B_r(x)\cap B_r(y) = \emptyset$. Then, the labels of $H_x$ and $H_y$ are computed with independent variables, so they will be independent. In general, with two sets $I$ and $J$ where $B_r(I)\cap B_r(J) = \emptyset$, given any $x\in I$ and $y\in J$, $H_x$ and $H_y$ will be computed with independent variables and so will be independent. So the output distribution is $r$-local.
\end{pf}

\paragraph{Quantum algorithms.}
When we discuss \emph{$r$-local quantum algorithms}, we are describing a quantum circuit with depth $r$ on $n$ qudits (each qudit separately measured in some basis $[d]$ for arbitrarily large $d$). The qudits are initialized in some product state. Each layer of the circuit can have single-qudit gates and a commuting set of  two-qudit gates. At the output level, we will partition $[d]$ into two disjoint parts $[d] = L \cup R$ and interpret each output qudit $x_i$ as corresponding to $+1$ if $x_i \in L$ and as corresponding to $-1$ if $x_i \in R$.

The QAOA of depth $p$ is an example of a $p$-local quantum algorithm. At each layer, there is the cost Hamiltonian $H_C = \sum_{e} H_e$ (where each $H_e$ corresponds to an edge $e$) and the mixing Hamiltonian $H_B = \sum_v H_v$ (where each $H_v$ corresponds to a vertex $v$). Within each Hamiltonian, the terms commute with each other, so the terms within each unitary also commute, corresponding to $U_C = \prod_e U_e$ and $U_B = \prod_v U_v$. So QAOA can be represented as a quantum circuit with local unitary gates, with each $U_v$ as a single-qubit gate and the $U_e$ as a set of commuting two-qubit gates. The qubits are then measured in the standard ($Z$) basis, with a qubit's output value referring to its partition.

\begin{thm}[Quantum local circuits induce local distributions]
\label{thm:quantumlocaldistribution}
Consider a $r$-local quantum algorithm for maximum cut. Given any graph $G$, the algorithm will output a cut from an $r$-local distribution.
\end{thm}
\begin{pf}
Consider the value of qudit $j$ with a one-local quantum algorithm. Let $O_j$ be the measurement operator on qudit $j$. With one layer of one-qudit and two-qudit gates, $O_j$ will become some $U^\dagger O_j U$, which is supported only on the neighborhood of $j$. So the distribution of the value of qudit $j$ depends only on $B_1(j)$. If another qudit $k$ is such that $B_1(j)\cap B_1(k) = \emptyset$, then the output values of $j$ and $k$ are independent; measuring qudit $j$ does not affect the measurement of qudit $k$. So the output distribution is one-local.

Now consider a $r$-local quantum algorithm. We iterate through each layer of the algorithm, working backwards from the measurement. At each layer, if the measurement operator is supported by vertices $S \subseteq V$ on the right-hand side of the layer, it can only be supported by $S'\subseteq B_1(S)$ on the left-hand side of the layer. So the output value of qudit $j$ depends only on $B_r(j)$; if $B_r(j)\cap B_r(k) = \emptyset$ for qudits $j,k$, then the output values of $j$ and $k$ are independent. So the output distribution is $r$-local.
\end{pf}

In a $r$-local quantum algorithm, the ``light cone'' $L_j$ of qudit $j$ refers to the set of input qudits that have a path through the circuit to the output qudit (i.e. $L_j \subseteq B_r(j)$). This is also defined for quantum algorithms with Hamiltonian terms that only act on a small number of qudits. Correlation bounds on quantum algorithms are often known as Lieb-Robinson bounds, where a quantitative relationship is drawn between parameters in the Hamiltonian and the evolution time \citep{liebrobinson, Pr_mont_Schwarz_2010}. Most often, the correlation with qudit $j$ decays exponentially with distance offset by $L_j$. Theorem \ref{thm:quantumlocaldistribution} is a strict kind of Lieb-Robinson bound for local quantum circuits, where there is zero correlation outside of the ``light cone'' (for example, see Lemma A.5 in \cite{lightcone} or Section 4.1 in \citet{farhi2020quantumTypicalCase}).

\paragraph{Centered distributions.} The maximum cut problem satisfies that the value of the cut $x$ is equal to the cut $-x$. Moreover, all natural classical or quantum randomized local algorithms we are aware of for this problem (and in particular QAOA) satisfy the symmetry property that for an individual vertex $i$, the probability that the output is $+1$ is the same as the probability that the output is $-1$. We will make this requirement from our distributions. We note that this is an extra assumption, but is a necessary one. 
For every graph $G_0$, there exists a $0$-local algorithm that outputs on all its inputs a fixed optimal cut  $x_0 \in \{ \pm 1 \}^n$ which is the optimum cut for this problem. Hence any pointwise lower bound along the lines of Theorem~\ref{thm:limit-informal} (which proves that \emph{all} $k$-local algorithms fail to achieve the optimum on certain graphs) must make an assumption such as centeredness. 

\end{document}